\documentclass[prd,preprintnumbers,floatfix,aps,nofootinbib,notitlepage,showpacs]{revtex4}
\usepackage{amsmath}
\usepackage{graphicx}
\usepackage{epstopdf}
\usepackage{latexsym}
\usepackage{amssymb}
\def\widebar{\overline}

\begin{document}

\title{A Review on the Cosmology of the de Sitter Horndeski Models}
\author{Nelson J. Nunes$^1$ and Prado Mart\'{\i}n-Moruno$^2$ and Francisco S. N. Lobo$^1$}
\affiliation{$^1$Instituto de Astrof\'isica e Ci\^encias do Espa\c{c}o, Universidade de Lisboa,
Faculdade de Ci\^encias, Campo~Grande, PT1749-016 Lisboa, Portugal \\
$^2$Departamento de F\'{\i}sica Te\'orica I, Universidad Complutense de Madrid, E-28040 Madrid
}

%\author{Nelson J. Nunes}
%\affiliation{Instituto de Astrof\'isica e Ci\^encias do Espa\c{c}o, Universidade de Lisboa,
%Faculdade de Ci\^encias, Campo~Grande, PT1749-016 Lisboa, Portugal}
%\author{Prado Mart\'{\i}n-Moruno}
%\affiliation{Departamento de F\'{\i}sica Te\'orica I, Universidad Complutense de Madrid, E-28040 Madrid}
%\author{Francisco S. N. Lobo}
%\affiliation{Instituto de Astrof\'isica e Ci\^encias do Espa\c{c}o, Universidade de Lisboa,
%Faculdade de Ci\^encias, Campo~Grande, PT1749-016 Lisboa, Portugal}

\begin{abstract}
We review the most general scalar-tensor cosmological models with up to second-order derivatives in the field equations that have a fixed spatially flat de Sitter critical point independent of the material content or vacuum energy. This subclass of the Horndeski Lagrangian is capable of dynamically adjusting any value of the vacuum energy of the matter fields at the critical point. We present the cosmological evolution of the linear models and the non-linear models with shift symmetry. We come to the conclusion that the shift symmetric non-linear models can deliver a viable background compatible with current observations.
\end{abstract}

\maketitle

%%%%%%%%%%%%%%%%%%%%%%%%%%%%%%%%%%%%%%%%%%

\section{Introduction}

The realisation that the Universe is currently undergoing an 
accelerated expansion is one of the major discoveries in cosmology. 
During the last eighteen years, a number of proposals to explain this 
evolution have been suggested. Most proposals involve scalar field 
dark energy (quintessence, k-essence, kinetic braiding) or extensions of Einstein's general relativity. 
These models are in principle stable, as their equations of motion are only second-order.
However, Lagrangians consisting of second-order derivatives generally give rise to equations of motion  with higher-order derivatives. Such theories might propagate a ghost degree of freedom, or in other words, they have an Ostrogradski instability \cite{Woodard:2006nt}.
In 1974, Horndeski wrote down the most general scalar-tensor theory leading to second-order equations of motion \cite{Horndeski:1974wa}. Despite being unnoticed for almost four decades, Deffayet~et~al.~\cite{Deffayet:2011gz} rediscovered this theory when generalizing the covariantized version \cite{Deffayet:2009wt} of the galileons models \cite{Nicolis:2008in}.  
It~turns out that Brans--Dicke theory, k-essence, kinetic braiding, or $f(R)$ models are subclasses of the most general Horndeski Lagrangian.
The theory can be written  in terms of the arbitrary functions $\kappa_i\left(\phi,\,X\right)$ and $F(\phi,\,X)$, where $X = \partial_\mu \phi \partial^\mu \phi$.
Thus, although the Horndeski theory restricts the type of stable scalar-tensor theories, there is still a huge amount of freedom.

As the vacuum energy  gravitates in extensions to general relativity, the cosmological constant problem  persists whenever the scalar field can only screen  a given value of that constant \cite{Weinberg:1988cp,Carroll:2000fy,Kaloper:2014dqa}.
In order to address this problem, Charmousis {et al.}~\cite{Charmousis:2011bf,Charmousis:2011ea} introduced the ``fab four'' models. In these models, the scalar field may acquire a non-trivial time dependence once the cosmological constant has been screened, hence avoiding Weinberg's no-go theorem. 
This screening  was constructed demanding that the critical point of the dynamics is Minkowski.
However, also by construction, as the dynamics approaches Minkowski, the universe is forced to decelerate. Therefore, a universe accelerating at late time does not naturally arise in this set up. 
 In this article, we review how the concept of self-adjustment was extended from Minkowskian to de Sitter final states  \cite{Martin-Moruno:2015bda} and show that  these models can lead to very promising cosmological scenarios from the observational point of view \cite{Martin-Moruno:2015lha,Martin-Moruno:2015eqa,Martin-Moruno:2015kaa}.

%%%%%%%%%%%%%%%%%%%%%%%%%%%%%%%%%%%%%%%
\subsection{Dynamical Screening}\label{general}

Let us consider a {FLRW} geometry of the universe.
After integrating the higher derivatives by parts, the Horndeski Lagrangian can be written as \cite{Charmousis:2011ea}
\begin{equation}\label{Lsimple}
 L(a,\dot a,\phi,\dot\phi)=a^3\sum_{i=0}^3 Z_i(a,\phi,\dot\phi)\,H^i,\qquad {\rm where}\qquad
 L=V^{-1}\int {\rm d}^3 x\,\mathcal{L}_H,
\end{equation}
$H=\dot a/a$ is the Hubble expansion rate, $V$ is the spatial integral of the volume element, and a dot identifies a derivative with respect to the cosmic time $t$. The functions $Z_i$ are written as
\begin{equation}\label{Z}
Z_i(a,\phi,\dot\phi)=X_i(\phi,\dot\phi)-\frac{k}{a^2} Y_i(\phi,\dot\phi),
\end{equation}
where $X_i$ and $Y_i$ are given in terms of the Horndeski free functions \cite{Charmousis:2011ea}. 
The Hamiltonian density~yields
\begin{equation}\label{Hamiltonian}
 \mathcal{H} (a,\dot a,\phi,\dot\phi)=\frac{1}{a^3}\left[\frac{\partial L}{\partial \dot{a}}\dot{a}+\frac{\partial L}{\partial \dot{\phi}}\dot{\phi}-L\right]=\sum_{i=0}^3\left[(i-1)Z_i+Z_{i,\dot\phi}\dot\phi\right]H^i.
\end{equation}
Let us assume that the matter fluids, given by the energy density $\rho_{\rm m}(a)$, are minimally coupled and do not interact with the scalar field. The Friedmann equation is then obtained from
\begin{equation}\label{MFE}
 \mathcal H + \mathcal H_{\rm EH} + \mathcal H_{\rm matter} = 0,
\end{equation}
where the Einstein--Hilbert Hamiltonian density is $\mathcal H_{\rm EH} = -3M_{\rm Pl}^2 H^2$ and the matter component is $\mathcal H_{\rm matter} = \rho_m$.
We will follow the same procedure described in Ref.~\cite{Charmousis:2011ea} applied to Minkwoski, but now requiring that self-tuning applies to a more general late-time solution  or critical point with $ H^2 \rightarrow H_c^2\neq0$.  Ideally, we would like this solution $H_c$  to be an attractor solution; however, this~particular adjustment mechanism can only ensure that it is a critical point.  The recipe for a~successful screening mechanism is the following:
\begin{enumerate}
 \item At the critical point, the field equation must be trivially satisfied such that the value of the scalar field is free to screen. This means that, up to a total derivative,  at the critical point the  Lagrangian density  must be independent of both $\phi$ and $\dot\phi$ 
 \begin{equation}\label{c1}
  \sum_{i=0}^3 Z_i(a_c,\phi,\dot\phi)H_c^i=c(a_c)+\frac{1}{a_c^3}\frac{d\zeta(a_c,\phi)}{dt}.
 \end{equation}
 This immediately shows that $\sum_i Z_i(a_c,\phi,\dot\phi)H_c^i$ is at most linear in $\dot\phi$.
 
\item In order to compensate for possible discontinuities of the cosmological constant appearing on the right hand side of the Friedmann equation, this equation must depend on $\dot\phi$ once screening has taken place. In other words, $\mathcal H_{,\dot\phi} \neq 0$. Taking into account Equation~(\ref{MFE}) and given that $\sum_i Z_{i,\dot\phi\dot\phi}(a_c,\phi,\dot\phi)H_c^i = 0$, as we saw above, it leads to
 \begin{equation}\label{c2}
  \sum_{i=1}^3i\,Z_{i,\dot\phi}(a_c,\phi,\dot\phi)H_c^i\neq0.
 \end{equation}

\item Requiring a non-trivial cosmology before screening implies that the scalar field equation of motion must depend on $\dot H$. This leads to the same condition (\ref{c2}) if $H_c\neq0$. In other words, $Z_{i,\dot\phi}(a_c,\phi,\dot\phi)H_c^i\neq0$ for at least one value of $i$. 

\end{enumerate}
Let us take a particular Lagrangian, $\widebar L$, that satisfies these  conditions at the critical point $a=a_c$, 
\begin{equation}
\widebar L =  \sum_{i=0}^3 \widebar Z_i(a_c,\phi,\dot\phi)H_c^i \,,
\end{equation}
and
\begin{equation}
 \sum_{i=1}^3i\,\widebar Z_{i,\dot\phi}(a_c,\phi,\dot\phi)H_c^i\neq0,
\end{equation}
as before, and where $\widebar Z_0$ is arbitrary. We now choose $\widebar Z_0$ such that at the critical point the Lagrangian is $\widebar L = c(a_c)$. The Lagrangian  is given quite generically as
 \begin{equation}\label{Lbarra}
 \widebar L(a,\dot a,\phi,\dot\phi)=a^3\left[c(a)+\sum_{i=1}^3\widebar Z_{i}(a,\phi,\dot\phi)\left(H^i-H_c^i\right)\right].
\end{equation}
By construction, this Lagrangian  has a critical point at $H_c$. 
We will now search for the form of the $\widebar Z_i$'s. As it was  explicitly shown in Ref.~\cite{Charmousis:2011ea}, 
two Horndeski Lagrangians which self-tune to $H_c$ are related by a total derivative of a function $\mu(a,\phi)$, such that
\begin{equation}\label{relacionL}
 L(a,\dot a,\phi,\dot\phi)=\widebar L (a,\dot a,\phi,\dot\phi)+\frac{d\mu(a,\phi)}{dt}.
\end{equation}
This relation must be valid during the whole evolution; therefore, equating equal powers of $H$, we~obtain
\begin{equation}\label{Zs}
 Z_0=c(a)-\sum_{i=1}^3\widebar Z_{i} H_c^i+\frac{\dot\phi}{a^3}\mu_{,\phi}, \qquad
 Z_1=\widebar Z_1+\frac{1}{a^2}\mu_{,a},\qquad Z_2=\widebar Z_2,\qquad Z_3=\widebar Z_3,
\end{equation}
which upon substituting $\widebar Z_i$ in the first of the above equations yields \cite{Charmousis:2011ea,Martin-Moruno:2015bda}
\begin{equation}\label{Zgral}
 \sum_{i=0}^3 Z_i(a,\phi,\dot\phi)H_c^i=c(a)+\frac{H_c}{a^2}\mu_{,a}(a,\phi)+\frac{\dot\phi}{a^3}\mu_{,\phi}(a,\phi).
\end{equation}

 %%%%%%%%%%%%%%%%%%%%%%%%%%%%%%%%%%%
 \subsection{The de Sitter Critical Point: $H_c^2=\Lambda$}\label{particular}

Let us first consider a flat universe with $k=0$,  which means that the dependence of  $Z_i$'s on 
the scale factor and $Y_i$'s disappears. We also require that $H_c^2=\Lambda$, which leads to
\begin{equation}
 \sum_{i=0}^3 X_i(\phi,\,\dot\phi)\Lambda^{i/2}=c(a)+\frac{\sqrt{\Lambda}}{a^2}\mu_{,a}+\frac{\dot\phi}{a^3}\mu_{,\phi}.
\end{equation}
As the left hand side of this equation is independent of $a$, so should the right hand side be for any value of $\dot\phi$. The function $\mu(a,\phi)$ must, therefore, be of the form
\begin{equation}
\mu(a,\phi) = a^3 h(\phi) - \frac{1}{\sqrt{\Lambda}} \int da \, c(a) a^2 .
\end{equation}
Thus, we have
\begin{equation}\label{Lonshell}
 L_c = a^3 \sum_{i=0}^3X_i(\phi,\dot\phi)\Lambda^{i/2}= a^3 \left(  3\sqrt{\Lambda}\,h(\phi)+\dot\phi\, h_{,\phi}(\phi) \right).
\end{equation}
Therefore, we can consider  three different kinds of terms in the Lagrangian. These are: (i) $X_i$-terms linear in $\dot\phi$;
(ii) $X_i$-terms non-linear in $\dot\phi$, the contribution of which must vanish at the critical point;
 and, (iii) terms not able to self-tune as they contribute via total derivatives, or terms that multiply by the curvature $k$ in the Lagrangian.

Let us now consider terms with an arbitrary dependence on $\phi$ and $\dot\phi$. The Lagrangian is given~by
\begin{equation}\label{Lnl}
 L=a^3 \sum_{i=0}^3 X_i(\phi,\dot\phi)\,H^i,
\end{equation}
and the Hamiltonian density  is
\begin{equation}\label{Hnlg}
 \mathcal{H} =\sum_{i=0}^{3}\left[(i-1)X_i(\phi,\dot\phi)+\dot\phi \,X_{i,\dot\phi}(\phi,\dot\phi)\right]H^i.
\end{equation}
The field equation can be written as 
\begin{eqnarray}\label{fieldnong}
 \sum_{i=0}^{3}   \left[X_{i,\phi}-3X_{i,\dot\phi}H- iX_{i,\dot\phi}\frac{\dot H}{H}-
 X_{i,\dot\phi \phi}\dot\phi-X_{i,\dot\phi\dot\phi}\ddot\phi\right] H^i = 0.
\end{eqnarray}

%%%%%%%%%%%%%%%%%%%%%%%%%%%%%%%%%%%%%%%%%%%%%%%%%%%%%%%%%%%%%%%%%%%%%%%
\section{Linear Models ``the Magnificent Seven''}

In order to satisfy Equation (\ref{Lonshell}) considering
only terms linear on $\dot\phi$, it is sufficient to set
\begin{equation}\label{lineargral}
 X_i^{\rm ms}(\phi,\dot\phi)=3\sqrt{\Lambda} \,U_i(\phi)+\dot\phi\,W_i(\phi),
\end{equation}
provided the potentials $U_i$ and $W_i$ satisfy the constraint
\begin{equation}\label{condnl}
  \sum_{i=0}^3W_i(\phi)\Lambda^{i/2}=\sum_{i=0}^3U_{i,\phi}(\phi)\Lambda^{i/2}.
\end{equation}
As there are in total eight functions $U_i$ and $W_i$, and only one constraint, there are seven free functions---the magnificent seven.
In these models,  the field equation  and the Friedmann equation read \cite{Martin-Moruno:2015lha},
\begin{eqnarray}
\label{H'}
 H'=3\frac{\sum_iH^i\left(\sqrt{\Lambda}\,  U_{i,\phi}(\phi)-H\,  W_{i}(\phi)\right)}{\sum_ii\,H^{i}  W_i(\phi)}, \qquad
 \label{phi'}
 \phi'=\sqrt{\Lambda}\frac{\left(1-\Omega\right)H^2-3\sum_i(i-1)\,H^i\,  U_i(\phi)}{\sum_ii\,H^{i+1}  W_i(\phi)},
\end{eqnarray}
where a prime means a derivative with respect to $\ln a$. The critical point of the system is $(H_c,\phi_c,\Omega) = (\sqrt{\Lambda},\phi_c,0)$, and its stability depends on the particular form of $U_i$ and $W_i$ \cite{Martin-Moruno:2015lha}. We are now going to consider a number of cases in our search for viable cosmological models compatible with current~observations. 

%%%%%%%%%%%%%%%%%%%%%%%%%%%%%%%%%%%%%%%%
\subsection{Only $W_0 \neq 0$}

Let us first assume that $W_0 \neq 0$, and $W_1=W_2=W_3=0$. In this case, $H'$ is ill-defined as the denominator of (\ref{H'}) vanishes. This can be understood by inspecting the Hamiltonian
\begin{eqnarray}
 \mathcal{H}_{\rm linear}&=&\sum_{i} \left[3(i-1)\sqrt{\Lambda}\,U_i(\phi)+i
\,\dot\phi\,W_i(\phi)\right]H^i = \sum_{i} \left[3(i-1)\sqrt{\Lambda}\,U_i(\phi)\right] H^i . \nonumber
\end{eqnarray}
We see that this Hamiltonian  is independent of $\dot\phi$, therefore violating condition (ii) for a successful Lagrangian. Thus, this model does not screen dynamically, and  only the de Sitter solution exists.

%%%%%%%%%%%%%%%%%%%%%%%%%%%%%%%%%%%%
\subsection{Only a $W_i$, $U_j$ Pair}

From the constraint equation, we have that $W_i = U_{j,\phi} \Lambda^{(j-i)/2}$, and then 
\begin{equation}
\frac{H'}{H} = -\frac{3}{i} \left[ 1-  \left( \frac{H}{\sqrt{\Lambda}} \right)^{j-i-1} \right], \nonumber
\end{equation}
which is independent  of $\phi$, and consequently, the matter content has no influence on the Universe's evolution. When $j-i-1<0$, the de Sitter solution is an attractor. When $H \gg \sqrt{\Lambda}$,  the field equation can be approximated  by $H'/H = -3/i$.
We can obtain a dust-like behaviour provided $i=2$ and---as~we expected by construction---we reach a de Sitter evolution when $H \rightarrow \sqrt{\Lambda}$.

%%%%%%%%%%%%%%%%%%%%%%%%%%%
\subsection{Only a $W_i$, $W_j$ Pair}

In this case, from the constraint equation,  $W_i = -W_{j,\phi} \Lambda^{(j-i)/2}$, and we have  
\begin{equation}
\small
\frac{H'}{H} = -3 \frac{1-(H/\sqrt{\Lambda})^{i-j}}{j-i(H/\sqrt{\Lambda})^{i-j}} , \nonumber
\end{equation}
which is again $\phi$ independent. 
For $j>i$, the de Sitter solution is an attractor. For $H \gg \sqrt{\Lambda}$, we can approximate the field equation as
$H'/H = -3/j$,
and  we recover a dust-like evolution for $j=2$.  A de Sitter universe is attained when $H \rightarrow \sqrt{\Lambda}$.

%%%%%%%%%%%%%%%%%%%%%%%%%%%%%%%%%%%%%%%%%%%%
\subsection{Term-by-Term Model}

We now consider that the constraint equation is satisfied for equal powers of $\Lambda$, such that $W_i = U_{i,\phi}$. There are eight functions and four constraints; therefore,  only four free potentials. 
Defining $U_{i,\phi} = \Lambda^{-i/2} V_{i,\phi}$, we can write
\begin{equation}
\small
\frac{H'}{H} = -3 \left( 1- \frac{\sqrt{\Lambda}}{H} \right)  \frac{\sum_i (H/
\sqrt{\Lambda})^i V_{i,\phi}}{\sum_i i (H/\sqrt{\Lambda})^i V_{i,\phi}}.\nonumber
\end{equation}
In this case, the scalar field contributes to the dynamics of
 the universe, as there is a dependence on $\phi$, which is itself determined by the matter content via Equation~(\ref{phi'}). 
For $H\gg \sqrt{\Lambda}$ and when only one $i$ component dominates, $H'/H = -3/i$, 
which means that dust is recovered for $i=2$. As before, we~reach de Sitter when $H \rightarrow \sqrt{\Lambda}$.

%%%%%%%%%%%%%%%%%%%%%%%%%%%%%%%%%%%%
\subsection{Tripod Model}

\begin{figure}[h]
\centering
\includegraphics[angle=0,width = 7cm]{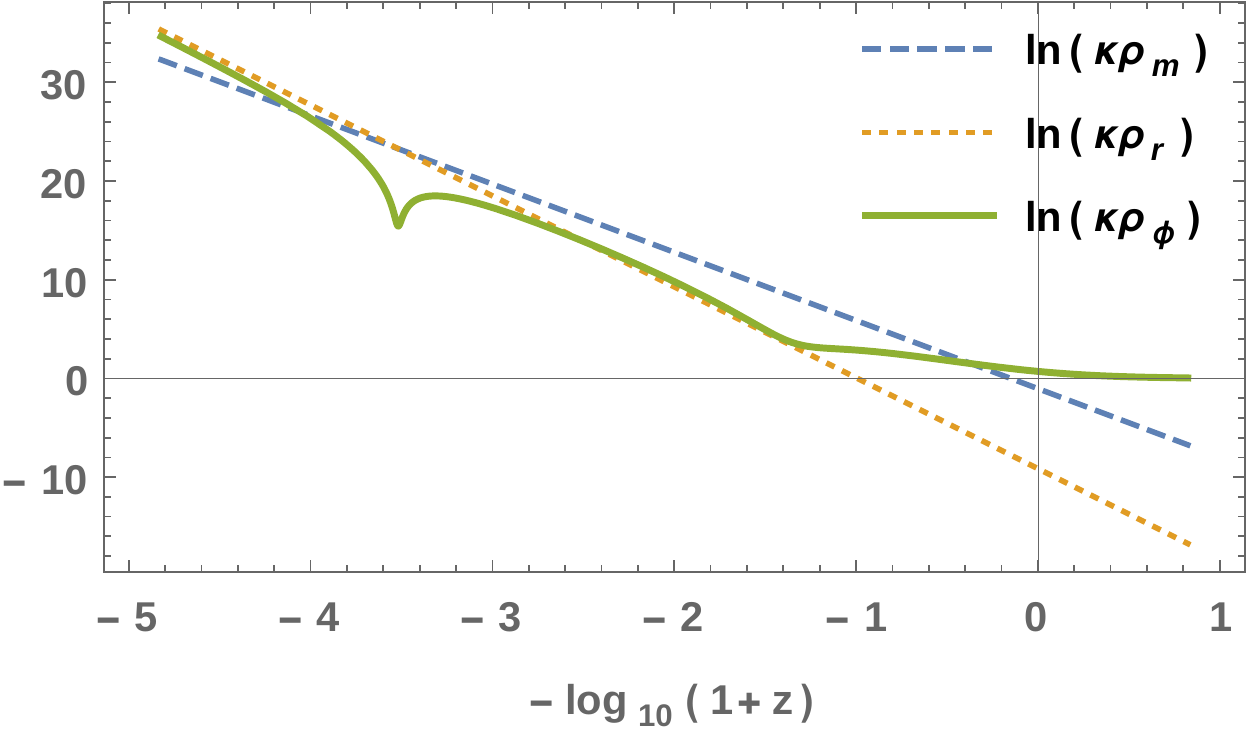} 
\caption{\label{tripod1} The evolution of the energy densities for the tripod models. Figure from \cite{Martin-Moruno:2015lha}.}

\end{figure}

Let us consider the three potentials $U_2$, $U_3$, and $W_2$. The constraint equation
 imposes $U_{2,\phi} \Lambda + U_{3,\phi} \Lambda^{3/2} = W_2 \Lambda$, and then 
\begin{equation}
\frac{H'}{H} = -3 \frac{U_{2,\phi}}{W_2} \left(1- \frac{\sqrt{\Lambda}}{H} \right).\nonumber
\end{equation}
For $H\gg \sqrt{\Lambda}$, we have approximately
\begin{equation}
\frac{H'}{H} = -\frac{3}{2}\frac{U_{2,\phi}}{W_2}. \nonumber
\end{equation}
In order to obtain a cosmological viable model,  we need:
$U_{2,\phi}/W_2 = 1$ during a matter domination epoch, and 
$U_{2,\phi}/W_2 = 4/3$ for a radiation domination epoch.
This can be achieved with the choice of potentials, $U_2 = e^{\lambda \phi} + \frac{4}{3} e^{\beta \phi}$, and
$W_2 = \lambda e^{\lambda \phi} + \beta e^{\beta \phi}$, as shown in Figure~\ref{tripod1}.
The de Sitter evolution is obtained when $H \rightarrow \sqrt{\Lambda}$.
Unfortunately, the field has a large contribution at early time which is incompatible with current~constraints.

%%%%%%%%%%%%%%%%%%%%%%%%%%%%%%%%%%%%%%%%
\section{Non-Linear Models}
In this section we consider that $X_i(\phi,\dot\phi)$ are  non-linear terms in $\dot\phi$ in the Lagrangian
\begin{equation}
 L_{\rm nl}=a^3 \sum_{i=0}^{3}X_i(\phi,\dot\phi)H^i.
 \end{equation}
As we saw before, any non-linear dependence of the Lagrangian on $\dot \phi$ must vanish at the 
critical point; thus, $\sum_{i=0}^{3} X_i(\phi,\dot\phi)\Lambda^{i/2}=0$. 
We will restrict the analysis to the  shift-symmetric cases, as it simplifies the calculations. Moreover, these cases also lead to a radiative stable situation since the field is non-renormalizable \cite{deRham:2012az}. 
Therefore, the system is independent of $\phi$, and we will make use of the convenient redefinition, $\psi = \dot\phi$. Under these assumptions, the field equation and the Friedmann equation are  \cite{Martin-Moruno:2015kaa}, 
\begin{eqnarray}
H'=\frac{3(1+w)Q_0P_1-Q_1P_0}{Q_1P_2-Q_2P_1}, \nonumber \hspace{1cm}
\psi'=\frac{3(1+w)Q_0P_2-Q_2P_0}{Q_2P_1-Q_1P_2},\nonumber
 \end{eqnarray}
where $Q_0$, $Q_1$, $Q_2$, $P_0$, $P_1$, $P_2$, are complicated functions of $X_i$ and $H$, 
and the average equation of state parameter of matter fluids is 
\begin{equation}
1+w=\frac{\sum_s\Omega_{s}(1+w_s)}{\sum_s\Omega_{s}}.\nonumber
 \end{equation}
The eigenvalues of the Jacobian matrix of the system formed by $H'$ and $\psi'$ evaluated at the critical point are  $(-3,-3(1+w))$, which means that the critical point  is stable whenever $w > -1$.

 As for the linear models, we are now going to take a systematic evaluation of the possible cosmological scenarios. In what follows, we will redefine $X_i$ such that 
 $ X_i = 3 M_{\rm Pl}^2 \Lambda^{1-i/2} f_i$.
 
 %%%%%%%%%%%%%%%%%%%%%%%%%%%%%%%%%%%%%%%%%
 \subsection{$f_3 = \psi^n$ Is the Dominant Contribution}

When $f_3$ is the dominant potential and $H\gg\sqrt{\Lambda}$, the effective equation of state is
\begin{eqnarray}
1+w_{\rm eff}&\simeq&\frac{2}{3}(1+w),\qquad{\rm for}\qquad \frac{|\left(2 f_3+\psi f_{3,\psi}\right)f_{3,\psi\psi}|}{|\left(3f_{3,\psi}+\psi f_{3,\psi\psi}\right)f_{3,\psi}|} \gg 1 \nonumber \\
1+w_{\rm eff} &\simeq&\frac{2}{3},  \qquad{\rm otherwise}. \nonumber
\end{eqnarray}
Neither of these allow for $w_{\rm eff}$ corresponding to radiation and/or matter domination epochs.

 %%%%%%%%%%%%%%%%%%%%%%%%%%%%%%%%%%%%%%%%%%%
\subsection{$f_2 = \psi^n$ Is the Dominant Contribution}

If instead $f_2$ is the dominant potential, for $H\gg\sqrt{\Lambda}$, it follows that  
\begin{eqnarray}
   w_{\rm eff} &\simeq& w,\qquad{\rm for}\qquad \frac{|\left(1-f_2-\psi f_{2,\psi}\right)f_{2,\psi\psi}|}{|\left(2f_{2,\psi}+\psi f_{2,\psi\psi}\right)f_{2,\psi}|} \gg 1, \nonumber \\ 
   w_{\rm eff} &\simeq& 0,\qquad{\rm otherwise}. \nonumber
\end{eqnarray}
In this case, either $w_{\rm eff}$ is too small at present when compared with observational constraints, or 
$\Omega_\psi$ is too large in the early universe.

%%%%%%%%%%%%%%%%%%%%%%%%%%%%%%%%%%%%%%%%%%%%%% 
 \subsection{$f_0$ and $f_1$ Are the Sole Contributions}

If we take $f_0$ and $f_1$ to be the only non-negligible potentials, then it can be shown that when \linebreak $H\gg\sqrt{\Lambda}$, the equation of state parameter  $w_{\rm eff}\simeq w$. This represents an
interesting case, but unfortunately, models with realistic initial conditions do not evolve to the critical point.

%%%%%%%%%%%%%%%%%%%%%%%%%%%%%%%%%%%%%%%%%
\subsection{Extension with $f_0$, $f_1$ and $f_2$}

\begin{figure}[h]
\centering
\includegraphics[width=0.5\textwidth]{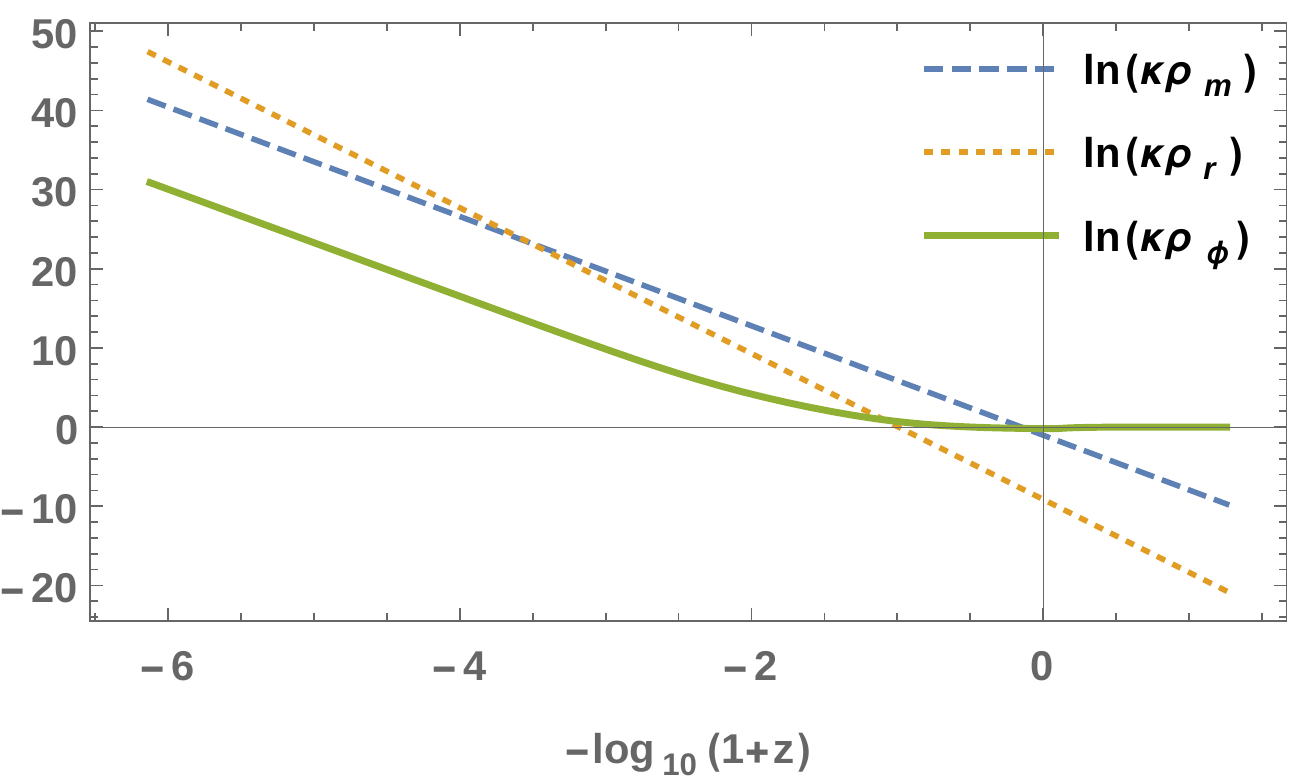}
\caption{\label{x0x1x2} The evolution of the energy densities for the model with non-negligible $X_0$, $X_1$, and $X_2$. Figure from \cite{Martin-Moruno:2015kaa}.}

\end{figure}

Finally, we consider a case involving the three potentials $X_0$, $X_1$, and $X_2$, such that
\begin{equation} 
f_2(\psi)=\alpha\psi^n,\qquad f_1(\psi)=-\alpha\psi^n+\frac{\beta}{\psi^m},\qquad f_0(\psi)=-\frac{\beta}{\psi^m}.\nonumber
\end{equation}
None of the potentials dominates the whole evolution; instead, different potentials are important at different epochs. 
This is a very promising case in what regards a background behaviour. We can obtain a model with $w_\psi = w_0 + w_a (1-a)$, such that $w_0 = -0.98$ and $w_a = 0.04$, which is compatible with current observational bounds. Moreover, the example gives  a negligible dark energy contribution at early times. The evolution of the energy densities of the field and matter fluids is illustrated in Figure~\ref{x0x1x2}.
 
 %%%%%%%%%%%%%%%%%%%%%%%%%%%%%%%%%%%
 \section{Summary}
 
 In this article, we have considered  a subclass of the Horndeski cosmological models that may alleviate the cosmological constant problem by screening any value the vacuum energy might take. They lead to a final de Sitter evolution of the universe regardless of the matter content. 
We have presented linear and the non-linear models and shown that the class of 
non-linear models with shift symmetry is very promising when tested against current observational constraints on the effective equation of state parameter and limits on early dark energy contribution. The natural following step of this work consists of investigating the evolution of linear perturbations of the field and of matter fluids in this scenario.

\acknowledgments{
This work was partially supported by the Funda\c{c}\~{a}o para a Ci\^{e}ncia e Tecnologia (FCT) through the grants EXPL/FIS-AST/1608/2013 and UID/FIS/04434/2013. PMM acknowledges financial support from the Spanish Ministry of Economy and Competitiveness (MINECO) through the postdoctoral training contract FPDI-2013-16161, and the project FIS2014-52837-P. 
NJN  was supported by a FCT Research contract, with reference IF/00852/2015. 
FSNL was supported by a FCT Research contract, with reference IF/00859/2012.}

\bibliography{references}

\begin{thebibliography}{15}
\expandafter\ifx\csname natexlab\endcsname\relax\def\natexlab#1{#1}\fi
\expandafter\ifx\csname bibnamefont\endcsname\relax
  \def\bibnamefont#1{#1}\fi
\expandafter\ifx\csname bibfnamefont\endcsname\relax
  \def\bibfnamefont#1{#1}\fi
\expandafter\ifx\csname citenamefont\endcsname\relax
  \def\citenamefont#1{#1}\fi
\expandafter\ifx\csname url\endcsname\relax
  \def\url#1{\texttt{#1}}\fi
\expandafter\ifx\csname urlprefix\endcsname\relax\def\urlprefix{URL }\fi
\providecommand{\bibinfo}[2]{#2}
\providecommand{\eprint}[2][]{\url{#2}}

\bibitem[{\citenamefont{Woodard}(2007)}]{Woodard:2006nt}
\bibinfo{author}{\bibfnamefont{R.~P.} \bibnamefont{Woodard}},
  \bibinfo{journal}{Lect. Notes Phys.} \textbf{\bibinfo{volume}{720}},
  \bibinfo{pages}{403} (\bibinfo{year}{2007}), \eprint{astro-ph/0601672}.

\bibitem[{\citenamefont{Horndeski}(1974)}]{Horndeski:1974wa}
\bibinfo{author}{\bibfnamefont{G.~W.} \bibnamefont{Horndeski}},
  \bibinfo{journal}{Int. J. Theor. Phys.} \textbf{\bibinfo{volume}{10}},
  \bibinfo{pages}{363} (\bibinfo{year}{1974}).

\bibitem[{\citenamefont{Deffayet et~al.}(2011)\citenamefont{Deffayet, Gao,
  Steer, and Zahariade}}]{Deffayet:2011gz}
\bibinfo{author}{\bibfnamefont{C.}~\bibnamefont{Deffayet}},
  \bibinfo{author}{\bibfnamefont{X.}~\bibnamefont{Gao}},
  \bibinfo{author}{\bibfnamefont{D.~A.} \bibnamefont{Steer}}, \bibnamefont{and}
  \bibinfo{author}{\bibfnamefont{G.}~\bibnamefont{Zahariade}},
  \bibinfo{journal}{Phys. Rev.} \textbf{\bibinfo{volume}{D84}},
  \bibinfo{pages}{064039} (\bibinfo{year}{2011}), \eprint{1103.3260}.

\bibitem[{\citenamefont{Deffayet et~al.}(2009)\citenamefont{Deffayet,
  Esposito-Farese, and Vikman}}]{Deffayet:2009wt}
\bibinfo{author}{\bibfnamefont{C.}~\bibnamefont{Deffayet}},
  \bibinfo{author}{\bibfnamefont{G.}~\bibnamefont{Esposito-Farese}},
  \bibnamefont{and} \bibinfo{author}{\bibfnamefont{A.}~\bibnamefont{Vikman}},
  \bibinfo{journal}{Phys. Rev.} \textbf{\bibinfo{volume}{D79}},
  \bibinfo{pages}{084003} (\bibinfo{year}{2009}), \eprint{0901.1314}.

\bibitem[{\citenamefont{Nicolis et~al.}(2009)\citenamefont{Nicolis, Rattazzi,
  and Trincherini}}]{Nicolis:2008in}
\bibinfo{author}{\bibfnamefont{A.}~\bibnamefont{Nicolis}},
  \bibinfo{author}{\bibfnamefont{R.}~\bibnamefont{Rattazzi}}, \bibnamefont{and}
  \bibinfo{author}{\bibfnamefont{E.}~\bibnamefont{Trincherini}},
  \bibinfo{journal}{Phys. Rev.} \textbf{\bibinfo{volume}{D79}},
  \bibinfo{pages}{064036} (\bibinfo{year}{2009}), \eprint{0811.2197}.

\bibitem[{\citenamefont{Weinberg}(1989)}]{Weinberg:1988cp}
\bibinfo{author}{\bibfnamefont{S.}~\bibnamefont{Weinberg}},
  \bibinfo{journal}{Rev. Mod. Phys.} \textbf{\bibinfo{volume}{61}},
  \bibinfo{pages}{1} (\bibinfo{year}{1989}).

\bibitem[{\citenamefont{Carroll}(2001)}]{Carroll:2000fy}
\bibinfo{author}{\bibfnamefont{S.~M.} \bibnamefont{Carroll}},
  \bibinfo{journal}{Living Rev. Rel.} \textbf{\bibinfo{volume}{4}},
  \bibinfo{pages}{1} (\bibinfo{year}{2001}), \eprint{astro-ph/0004075}.

\bibitem[{\citenamefont{Kaloper and Padilla}(2014)}]{Kaloper:2014dqa}
\bibinfo{author}{\bibfnamefont{N.}~\bibnamefont{Kaloper}} \bibnamefont{and}
  \bibinfo{author}{\bibfnamefont{A.}~\bibnamefont{Padilla}},
  \bibinfo{journal}{Phys. Rev.} \textbf{\bibinfo{volume}{D90}},
  \bibinfo{pages}{084023} (\bibinfo{year}{2014}), \bibinfo{note}{[Addendum:
  Phys. Rev.D90,no.10,109901(2014)]}, \eprint{1406.0711}.

\bibitem[{\citenamefont{Charmousis
  et~al.}(2012{\natexlab{a}})\citenamefont{Charmousis, Copeland, Padilla, and
  Saffin}}]{Charmousis:2011bf}
\bibinfo{author}{\bibfnamefont{C.}~\bibnamefont{Charmousis}},
  \bibinfo{author}{\bibfnamefont{E.~J.} \bibnamefont{Copeland}},
  \bibinfo{author}{\bibfnamefont{A.}~\bibnamefont{Padilla}}, \bibnamefont{and}
  \bibinfo{author}{\bibfnamefont{P.~M.} \bibnamefont{Saffin}},
  \bibinfo{journal}{Phys. Rev. Lett.} \textbf{\bibinfo{volume}{108}},
  \bibinfo{pages}{051101} (\bibinfo{year}{2012}{\natexlab{a}}),
  \eprint{1106.2000}.

\bibitem[{\citenamefont{Charmousis
  et~al.}(2012{\natexlab{b}})\citenamefont{Charmousis, Copeland, Padilla, and
  Saffin}}]{Charmousis:2011ea}
\bibinfo{author}{\bibfnamefont{C.}~\bibnamefont{Charmousis}},
  \bibinfo{author}{\bibfnamefont{E.~J.} \bibnamefont{Copeland}},
  \bibinfo{author}{\bibfnamefont{A.}~\bibnamefont{Padilla}}, \bibnamefont{and}
  \bibinfo{author}{\bibfnamefont{P.~M.} \bibnamefont{Saffin}},
  \bibinfo{journal}{Phys. Rev.} \textbf{\bibinfo{volume}{D85}},
  \bibinfo{pages}{104040} (\bibinfo{year}{2012}{\natexlab{b}}),
  \eprint{1112.4866}.

\bibitem[{\citenamefont{Martin-Moruno et~al.}(2015)\citenamefont{Martin-Moruno,
  Nunes, and Lobo}}]{Martin-Moruno:2015bda}
\bibinfo{author}{\bibfnamefont{P.}~\bibnamefont{Martin-Moruno}},
  \bibinfo{author}{\bibfnamefont{N.~J.} \bibnamefont{Nunes}}, \bibnamefont{and}
  \bibinfo{author}{\bibfnamefont{F.~S.~N.} \bibnamefont{Lobo}},
  \bibinfo{journal}{Phys. Rev.} \textbf{\bibinfo{volume}{D91}},
  \bibinfo{pages}{084029} (\bibinfo{year}{2015}), \eprint{1502.03236}.

\bibitem[{\citenamefont{Martín-Moruno
  et~al.}(2015)\citenamefont{Martín-Moruno, Nunes, and
  Lobo}}]{Martin-Moruno:2015lha}
\bibinfo{author}{\bibfnamefont{P.}~\bibnamefont{Martín-Moruno}},
  \bibinfo{author}{\bibfnamefont{N.~J.} \bibnamefont{Nunes}}, \bibnamefont{and}
  \bibinfo{author}{\bibfnamefont{F.~S.~N.} \bibnamefont{Lobo}},
  \bibinfo{journal}{JCAP} \textbf{\bibinfo{volume}{1505}}, \bibinfo{pages}{033}
  (\bibinfo{year}{2015}), \eprint{1502.05878}.

\bibitem[{\citenamefont{Martin-Moruno and
  Nunes}(2015{\natexlab{a}})}]{Martin-Moruno:2015eqa}
\bibinfo{author}{\bibfnamefont{P.}~\bibnamefont{Martin-Moruno}}
  \bibnamefont{and} \bibinfo{author}{\bibfnamefont{N.~J.} \bibnamefont{Nunes}},
  \bibinfo{journal}{Int. J. Mod. Phys.} \textbf{\bibinfo{volume}{D24}},
  \bibinfo{pages}{1544018} (\bibinfo{year}{2015}{\natexlab{a}}),
  \eprint{1505.06585}.

\bibitem[{\citenamefont{Martin-Moruno and
  Nunes}(2015{\natexlab{b}})}]{Martin-Moruno:2015kaa}
\bibinfo{author}{\bibfnamefont{P.}~\bibnamefont{Martin-Moruno}}
  \bibnamefont{and} \bibinfo{author}{\bibfnamefont{N.~J.} \bibnamefont{Nunes}},
  \bibinfo{journal}{JCAP} \textbf{\bibinfo{volume}{1509}}, \bibinfo{pages}{056}
  (\bibinfo{year}{2015}{\natexlab{b}}), \eprint{1506.02497}.

\bibitem[{\citenamefont{de~Rham}(2012)}]{deRham:2012az}
\bibinfo{author}{\bibfnamefont{C.}~\bibnamefont{de~Rham}},
  \bibinfo{journal}{Comptes Rendus Physique} \textbf{\bibinfo{volume}{13}},
  \bibinfo{pages}{666} (\bibinfo{year}{2012}), \eprint{1204.5492}.

\end{thebibliography}

\end{document}